# Inter-channel Conv-TasNet for multichannel speech enhancement

Dongheon Lee, Seongrae Kim, *Student Member, IEEE*, and Jung-Woo Choi, *Member, IEEE*

*Abstract*—Speech enhancement in multichannel settings has been realized by utilizing the spatial information embedded in multiple microphone signals. Moreover, deep neural networks (DNNs) have been recently advanced in this field; however, studies on the efficient multichannel network structure fully exploiting spatial information and inter-channel relationships is still in its early stages. In this study, we propose an end-to-end time-domain speech enhancement network that can facilitate the use of inter-channel relationships at individual layers of a DNN. The proposed technique is based on a fully convolutional time-domain audio separation network (Conv-TasNet), originally developed for speech separation tasks. We extend Conv-TasNet into several forms that can handle multichannel input signals and learn inter-channel relationships. To this end, we modify the encoder-mask-decoder structures of the network to be compatible with 3-D tensors defined over spatial channels, features, and time dimensions. In particular, we conduct extensive parameter analyses on the convolution structure and propose independent assignment of the depthwise and 1×1 convolution layers to the feature and spatial dimensions, respectively. We demonstrate that the enriched inter-channel information from the proposed network plays a significant role in suppressing noisy signals impinging from various directions. The proposed inter-channel Conv-TasNet outperforms the state-of-the-art multichannel variants of neural networks, even with one-tenth of their parameter size. The performance of the proposed model is evaluated using the CHiME-3 dataset, which exhibits a remarkable improvement in SDR, PESQ, and STOI.

## I. Introduction

Multichannel speech enhancement is a task to recover clean target speech from noise mixtures recorded by multichannel microphones. In the past, clean speech was estimated using various spatial filters designed using beamforming techniques, such as EB-MVDR [1] and EB-MUSIC [2]. In recent years, deep learning (DL)-based speech enhancement techniques have exhibited acceptable performance in single-channel speech enhancement tasks [3]–[5]. The DL-based speech enhancement technique directly estimates the target sound source with a reduced noise through nonlinear mapping operations realized by linear transform and nonlinear activation units. Deep neural networks (DNNs) are incorporated in the DL-based approach, which can model complex nonlinear relationships.

The DL-based speech enhancement technique produces a clean speech signal by estimating a spectrogram or time-domain waveform [6], [7]. In the spectrogram estimation approach, a DNN is configured to learn the mapping between the input and output spectrograms. The spectrogram handled by the DNN can be either a magnitude or a complex spectrogram. When only the magnitude spectrogram is used, the noisy phase information can be combined with the estimated magnitude spectrogram [6], [8], [9], or the clean phase information can be reconstructed using a separate network or post-processing technique [10]. Furthermore, DNNs using the complex spectrogram as input and output features [11] have been proposed to mitigate the phase reconstruction issue. However, both spectrogram-based approaches require the short-time Fourier transform (STFT) [12] for spectrogram conversion, which is not optimized for speech enhancement tasks. Therefore, end-to-end approaches have been proposed to directly receive and generate time-domain signals [7], [13]. Training and inference using temporal waveforms automatically consider both the magnitude and phase information and enable the learning of appropriate characteristics for noise reduction.

The actual speech enhancement can be realized using the direct estimation of a target speech signal or noise suppression mask estimation. In the latter approach, a real or complex mask is trained by a DNN such that the element-wise multiplication of a noisy speech with a mask function reconstructs a clean speech [14]–[16]. The mask estimation approach originates from the ideal binary mask of computational auditory scene analysis [17], which was subsequently extended to the ideal ratio mask [18], [19]. In recent years, many DNNs have been introduced for mask estimation [20], [21].

The success of the single-channel DNN in the speech enhancement task has evolved into multichannel DNN models. Unlike the single-channel case, more spatial information is available for multichannel speech enhancement tasks. Therefore, the extraction of spatial features is key to the success of multichannel speech enhancement tasks. Many neural beamforming techniques have been proposed that use beamforming as the front end of a single-channel neural network [22]–[24]. Neural beamforming techniques were designed to combine the benefits of spatial filtering and DL;

This work was supported by the BK21 Four program through the National Research Foundation (NRF) funded by the Ministry of Education of Korea and conducted by Center for Applied Research in Artificial Intelligence (CARAI) grant funded by DAPA and ADD (UD190031RD). (*Corresponding author: Jung-Woo Choi.*) The authors are with the School of Electrical Engineering, Korea Advanced Institute of Science and Technology (KAIST), Daejeon 34141, South Korea (e-mail: dongheon0115@kaist.ac.kr; guana@kaist.ac.kr; jwoo@kaist.ac.kr).

however, a fundamental performance limit set exists in the linear spatial filtering of the beamforming process.

Therefore, many DNN-based techniques have been developed that can exploit nonlinear relationships between microphones and overcome this limit [25], [26]. To date, Channel-Attention Dense U-Net (CA Dense U-Net) [27] has exhibited the best performance among the models tested using the CHiME-3 dataset [28]. CA Dense U-Net introduced the channel-attention network into the conventional U-Net structure to mimic beamforming. Despite its high performance, several issues [21] arise in the STFT procedure and U-Net [29] structure. First, it is uncertain whether the Fourier transform is the optimal transformation candidate for a speech enhancement task. Second, because STFT converts the time-domain signal into a complex domain, the network needs to handle both the magnitude and phase of the signal. Although a complex ratio mask has been proposed to utilize the phase information [11], the upper bound of performance exists because the reconstruction artifact induced by the up-sampling process. Third, the U-Net-based network typically has a large parameter size. Thus, the signal should be directly modeled in the time-domain using structures such as TasNet [21] to overcome these problems.

Among DNN models employing the TasNet structure, Conv-TasNet [30] is an advanced end-to-end architecture appropriate for speech separation tasks. Conv-TasNet utilizes a temporal convolutional network (TCN) [31], [32] consisting of dilated one-dimensional convolution (1-D Conv) layers. This can reduce the checkerboard artifact induced by down- and up-sampling blocks of U-Net structures, which might be beneficial to the speech enhancement task. Its computation time can be reduced by performing parallel convolution, which allows its extension to a larger-scale model, such as the multichannel structure presented in this study.

Following the success of Conv-TasNet in single-channel speech enhancement [33], [34], its extensions to the multichannel speech enhancement task [35], [36] have been attempted. One representative example is the multichannel Conv-TasNet (MC Conv-TasNet) [35], which has a multichannel encoder to handle multichannel input data, and adds up encoded multichannel features into a single-channel. However, the spatial information is lost by the addition; thus, it cannot fully utilize the multichannel information in the TCN layers.

In this study, we propose a multichannel DNN model that can enhance noisy speech by fully exploiting the inter-channel relationship in the multichannel data. The base architecture of the proposed model is the MC Conv-TasNet; however, the encoder-mask-decoder structure of the conventional network is substantially redesigned to extract the spatial relationship using the network.

The proposed model, referred to as inter-channel Conv-TasNet (IC Conv-TasNet) in this study, has the following advanced characteristics:

1. We constitute a 3-D tensor by stacking multichannel encoder outputs in the channel dimension. This changes the 2-D input feature of MC Conv-TasNet to a 3-D feature with various spatial information.

2. We separate the roles of the depthwise and 1-D convolutions of the TCN such that the depthwise convolution extracts inter-channel relationships only, whereas the 1-D Conv layer focuses on the extraction of spectral and temporal features. To this end, the 1-D Conv layer is replaced by a 2-D Conv functioning in the feature and time dimensions.

Additionally, an extensive parameter study was conducted to build a small-sized DNN model. The model with the best performance was determined using various parameter studies with different model parameter sizes. The superior performance of the proposed model is compared with that of the state-of-the-art (SOTA) models [27], [37] developed for multichannel speech enhancement tasks.

The remainder of this paper is organized as follows. Section II introduces the proposed model architecture and provides details of the IC Conv-TasNet. In Sections III and IV, performance evaluation is presented with parameter studies. A downsized version of the IC Conv-TasNet is also introduced using a detailed analysis of the parameter size. Finally, Section V summarizes the results and draws the conclusions.

## II. MODEL ARCHITECTURE

The proposed DNN model is based on Conv-TasNet. The conventional single-channel (SC) Conv-TasNet for the sound separation task consists of an encoder, mask(separation), and decoder structures. The input waveform is divided into $L$ overlapping segments of window length $K$, which are represented as $\mathbf{x} \in \mathbb{R}^{L \times K}$. The 1-D Conv layer in the encoder module converts each segment data into a feature vector of length $F$. The encoder output $\mathbf{w} \in \mathbb{R}^{L \times F}$ can be written as

$$\mathbf{w} = \text{ReLU}(\mathbf{x}U), \qquad (1)$$

where $U \in \mathbb{R}^{K \times F}$ is the encoding matrix, and ReLU [38] is the rectified linear unit to guarantee the nonnegative representation of $\mathbf{w}$.

The encoder output is subsequently fed into the mask estimation network (separation module) and analyzed to estimate the source separation mask. The separation of the $i^{th}$ source ($i = 1,...,S$) is realized by the element-wise product of the estimated mask $\mathbf{m}_i \in \mathbb{R}^{L \times F}$ and the encoder output $\mathbf{w}$. The masked encoder outputs of the $i^{th}$ source can be obtained as

$$\mathbf{d}_i = \mathbf{w} \odot \mathbf{m}_i, \qquad (2)$$

where $\odot$ denotes the element-wise product. In the sound enhancement task, the source index $i$ is discarded, and we obtain $\mathbf{d} = \mathbf{w} \odot \mathbf{m}$.

A decoder subsequently takes the masked encoder outputs and reconstructs them into enhanced sound segments. The segments of the enhanced sound can be expressed as

$$\hat{\mathbf{s}} = \mathbf{d}V, \qquad (3)$$

where $V \in \mathbb{R}^{F \times K}$ is the decoding matrix. The reconstructed segments can be transformed into the final waveform using an overlap-and-add operation [39]. The overall structure of Conv-TasNet is presented in Fig. 1.

The source separation module of Conv-TasNet consists of a single 1×1 Conv layer compressing the number of features from $F$ to $N$ ($F > N$), followed by several stacks of 1-D Conv blocks. Here, we denote the 1×1 Conv layer before the stacks of 1-D Conv blocks as a 'bottleneck layer.' Each 1-D Conv block (Fig. 2(a)) includes a 1×1 Conv layer upsizing the number of features from $N$ to the number of hidden layers $H$, dilated depthwise convolution (D-Conv) layer [40], [41], and two 1×1 Conv layers for the residual path and skip connection, respectively. Each 1-D Conv block implements its own expansion and compression of features before and after the D-Conv. The nonlinear activation function (PReLU [42]) and layer normalization are applied after 1×1 Conv and D-Conv, respectively. The serial connection of 1-D Conv blocks constituting a single TCN stack increases the receptive field by using a higher dilation factor $d$. In each stack of TCN, $D$ 1-D Conv blocks are located. Multiple stacks are used to extract different information from multiple skip connections, which are subsequently added and mixed to generate a source separation mask. The generated mask is multiplied with the single-channel encoder output to produce a separated waveform through the transposed convolution layer.

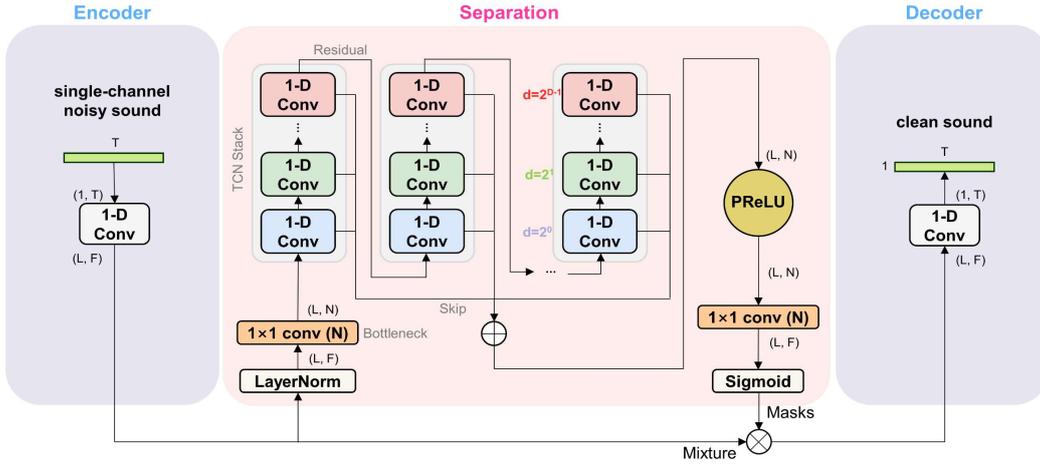

Fig. 1. Overall structure of Conv-TasNet

MC Conv-TasNet [35] incorporated one modification into SC Conv-TasNet to handle multichannel data. The encoder is extended in the channel dimension and 1-D Conv operations are applied to individual microphone channels. The 3-D tensor output of size (time, feature, channel) = $(L, F, M)$, obtained from individual encoders is then superposed along the microphone channel dimension to produce a 2-D encoder output feature of $\mathbf{w} \in \mathbb{R}^{L \times F}$, which is compatible with other layers of the SC Conv-TasNet, as shown in Fig. 3(a). However, this superposition of multichannel output causes the inter-channel relationship to be lost in the following convolutional layers.

Here, we introduce and compare three different models using a modified encoder and TCN structures to seek a network structure that can effectively extract inter-channel relationship throughout all convolution layers.

### A. 2-D Conv-TasNet

To avoid the loss of spatial information accompanied by the addition operation of the MC Conv-TasNet, we concatenate the multichannel encoder outputs, which yields a 2-D tensor $\mathbf{w} \in \mathbb{R}^{L \times FM}$. Additionally, the bottleneck layer is modified to downsize the number of features ($FM$) to the number of features $N$. Consequently, the channel and feature information are mixed by the 1×1 Conv layer before TCN. Because the rest of the architecture is the same as that of Conv-TasNet, no difference exists between the channel dimension $M$ and feature dimension $F$ in TCN.

We denote this network with a modified encoder and bottleneck as 2-D Conv-TasNet. The block diagrams of 1-D Conv block in TCN and the encoder module are shown in Figs. 2(a) and 3(b), respectively. One extra modification required for the multichannel structure is the mask multiplication step. As the encoder outputs $\mathbf{w} \in \mathbb{R}^{L \times FM}$ are multichannel data, we select one microphone channel feature $\mathbf{w}_{ref} = \mathbb{R}^{L \times F}$ among $\mathbf{w}$ and multiply the generated mask to that channel only. The selected channel is indicated as the reference encoder output in this study.

### B. 3-D Conv-TasNet

In the next architecture, indicated as 3-D Conv-TasNet, the channel dimension is separately handled from the feature dimension, as presented in Fig. 3(c). A 3-D tensor of size ($L$, $F$, $M$) is constructed by stacking the outputs of the multichannel encoder and changing its size to ($L$, $N$, $C$) through two 1×1 Conv layers positioned before TCN. By constructing a 3-D tensor and applying separate 1×1 Conv operations along the channel and feature dimensions, we attempt to independently treat the channel and feature information throughout the entire TCN layer.

The 3-D tensor is subsequently split into $C$ slices of ($L$, $N$) matrices, which are sent to the $C$ individual 1-D Conv blocks of TCN, as shown in Fig. 2(b). Consequently, different channels are independently convolved in individual TCN layers, and the mix of channel information occurs only at the final mask

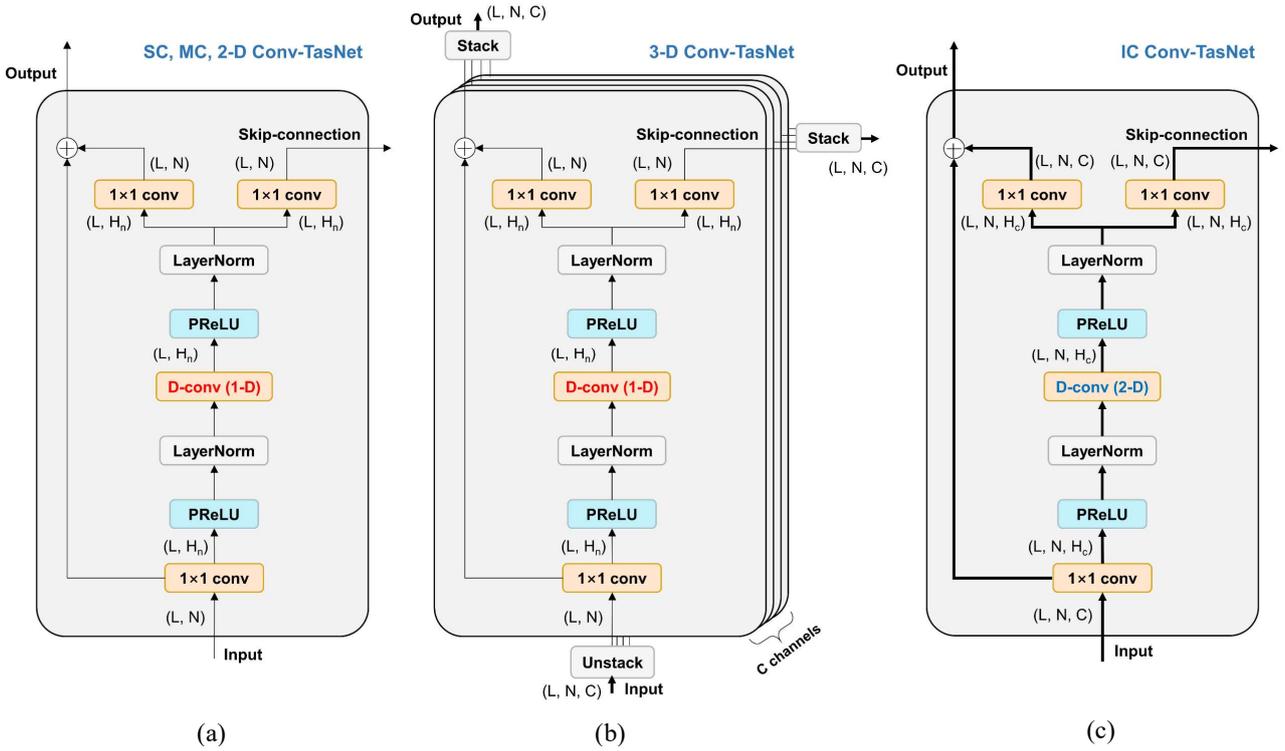

Fig. 2. Structures of 1-D Conv blocks in TCN of (a) SC Conv-TasNet, MC Conv-TasNet and 2-D Conv-TasNet (b) 3-D Conv-TasNet, and 2-D Conv block of (c) IC Conv-TasNet

generation step by two 1×1 Conv layers positioned after the parametric rectified linear unit (PReLU).

*C. Inter-channel Conv-TasNet (IC Conv-TasNet)*

IC Conv-TasNet is the final model proposed in this study, as illustrated in Fig. 4. Unlike the 3-D Conv-TasNet that utilizes the inter-channel relationship at the mask generation stage, the IC Conv-TasNet extracts inter-channel features within the TCN layers to fully exploit the available spatial information.

Similar to 3-D Conv-TasNet, 1-D Conv is applied to each microphone channel in the encoder module, and the encoder outputs of individual channels are stacked in the channel dimension, as shown in Fig. 3(c). The major difference between IC Conv-TasNet is the TCN blocks of the mask estimation network.

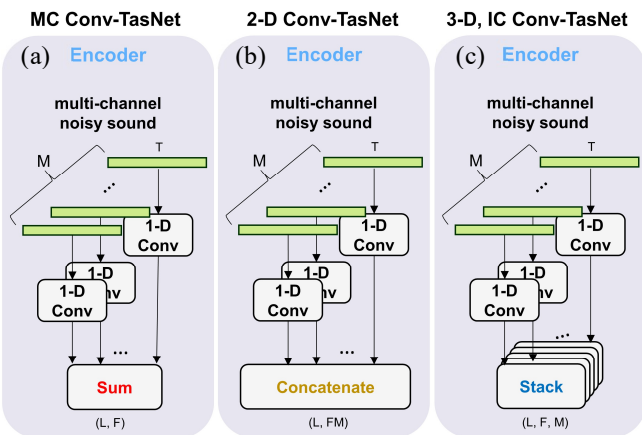

Fig. 3. Comparison of encoder structures of (a) MC Conv-TasNet (b) 2-D Conv-Conv-TasNet (c) 3-D and IC Conv-TasNet

In the 2-D Conv block in the TCN of IC Conv-TasNet (Fig. 2(c)), only the size of the channel dimension is increased by 1×1 Conv from $C$ to the number of hidden layers $H$. This modification promotes channel-wise diversity without changing feature and time dimensions. The increased channel size is set as four times the input channel size in this study ($H = 4C$).

Subsequently, the D-Conv layer is modified to apply 2-D depthwise convolutions in the feature and time dimensions. The D-Conv layer receives a feature map of size ($L$, $N$, $C$) as an input, and extracts feature- and time-dependent information from 2-D dilated convolution of $C$ slices of ($L$, $N$) matrices. Furthermore, 2-D zero padding is employed in the D-Conv layer to maintain the size of the feature map.

The inter-channel relationships are then extracted by two 1×1 Conv layers located before the skip and residual paths. Unlike the other Conv-TasNets that apply a 1×1 convolution in the feature dimension, the 1×1 Conv layers of the IC Conv-TasNet only extract the information in the channel dimension to focus on inter-channel relationships.

After the skip connection is connected to the PReLU activation function, the channel information is compressed through a 1×1 Conv ($C$) layer to make the feature map into a single-channel. Subsequently, another 1×1 Conv ($N$) layer adjusts the number of features from $N$ to that of the encoder output $F$. A sigmoid function is used to constrain the values of the mask within [0, 1]. The generated mask of size ($L$, $F$) is multiplied with the encoder output in terms of an element-wise product, resulting in the masked encoder outputs.

performance and parameter size of the best model were also compared to those of the SOTA models designed for multichannel speech enhancement. Finally, the third

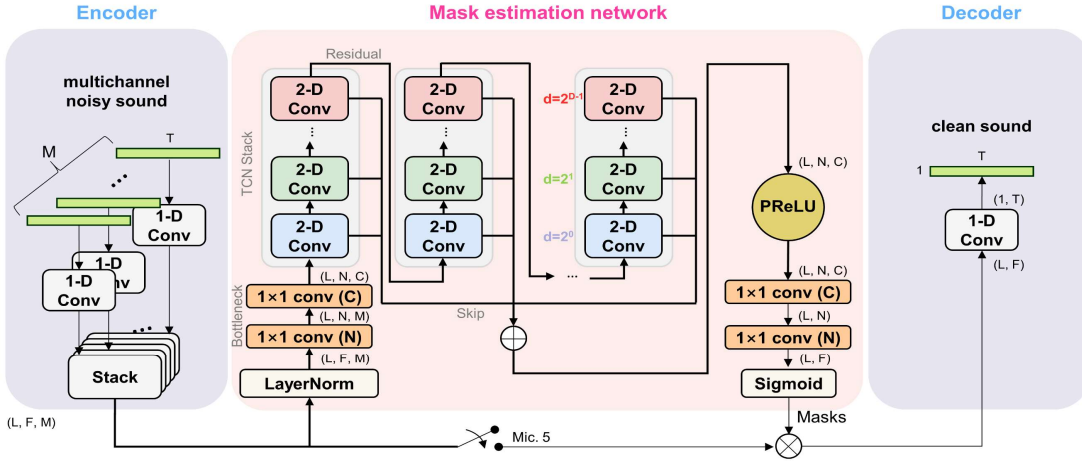

Fig. 4. Overall structure of IC Conv-TasNet

In summary, the 2-D Conv block in TCN aggregates the information of each channel through 1×1 Conv; thus, each Conv layer actively exploits inter-channel relationship to extract features. The entire model architecture of the IC Conv-TasNet and 2-D Conv block in TCN are illustrated in Figs. 4 and 2(c).

### III. EXPERIMENT

#### A. Dataset

The three models (2-D, 3-D, and IC Conv-TasNet) discussed in Section II were tested using the publicly available CHiME-3 dataset [28]. The dataset was designed for speech enhancement and recognition tasks. Moreover, it has been commonly used for testing multichannel speech enhancement techniques [25], [26]. The dataset was created by mixing a speech pronounced using a speaker positioned at different locations from the tablet device and outdoor noise. The outdoor noises were recorded in four different outdoor environments, including cafés, buses, pedestrian areas, and streets. The speeches were measured using 6-channel microphones installed on the tablet device at a sampling rate of 16 kHz. This dataset contains 7,138, 1,640, and 1,320 simulated utterances, which correspond to the training, development, and test data, respectively.

#### B. Experiment procedure

We used the MC Conv-TasNet as the baseline model for the performance comparison. In the first experiment was conducted to compare the performance of the three variants of Conv-TasNet. We chose the best model architecture from a comparison to the baseline. The second experiment was designed to search for the best model parameters through various parameter studies. The parameter studies were conducted by changing the number of layers in each stack of TCN ($D$), number of TCN stacks ($S$), number of features ($F$, $N$), and size of channel dimension ($C$) as listed in Table 1. The experiment was designed to explore the possibility of downsizing the IC Conv-TasNet. By designing a downsized model with a reduced input tensor size, we attempted to reduce the training time with a minimal effect on the performance.

Additionally, the networks were trained in the experiments using the signal-to-distortion ratio (SDR) loss for 200 epochs. The encoder and decoder processed the temporal waveform using a window with a length of 256 and overlap of 50%. A 3×3 kernel size was used for the 1-D or 2-D Conv block of TCN. Channel 5 was selected as the reference encoder output, based on suggestions from a previous study [43]. The network was trained with an ADAM [44] optimizer with a learning rate of $10^{-3}$.

TABLE 1
HYPERPARAMETERS OF THE NETWORK

| Symbol | Description |
| --- | --- |
| $D$ | Number of dilated Conv layers in each stack |
| $S$ | Number of stacks in TCN |
| $F$ | Number of encoder outputs |
| $N$ | Number of kernels in the bottleneck layer |
| $C$ | Channel dimension |

#### C. Evaluation metrics

The performance of the proposed models was evaluated in terms of three metrics typically used in the speech enhancement task: SDR [45], perceptual evaluation of speech quality (PESQ; the wideband version recommended in ITU-T P.862.2 [46], and the short-time objective intelligibility measure (STOI) [47]. A short description of these metrics is presented below:

1) **Signal-to-Distortion Ratio (SDR)**
SDR is an objective metric designed to express the physical quality of the reconstructed signal. That is, SDR represents the distance between the ground truth and recovered signals, which

TABLE 2
COMPARISON OF MODEL ARCHITECTURES

|  | Hyper parameters | | | | | | Param. Size | Performance indices | | |
| --- | --- | --- | --- | --- | --- | --- | --- | --- | --- | --- |
|  | $D$ | $S$ | $F$ | $N$ | $C$ | $H$ |  | SDR | PESQ | STOI |
| MC Conv-TasNet | 8 | 3 | 2048 | 512 | 1 | 2048 | 79.1 M | 13.36 | 1.58 | 0.912 |
| 2-D Conv-TasNet | 8 | 3 | 2048 | 512 | 1 | 2048 | 84.4 M | 15.58 | 1.84 | 0.940 |
| 3-D Conv-TasNet | 8 | 3 | 2048 | 64 | 8 | 32 | 2.56 M | 15.71 | 2.05 | 0.945 |
| **IC Conv-TasNet** | **8** | **3** | **2048** | **64** | **8** | **32** | **1.35 M** | **16.51** | **2.24** | **0.949** |
| **IC Conv-TasNet (best performance)** | **8** | **3** | **512** | **128** | **64** | **256** | **1.67 M** | **19.67** | **2.67** | **0.973** |

is defined as follows:

$$\text{SDR}(\mathbf{s}, \hat{\mathbf{s}}) = 20\log_{10}\frac{\|\mathbf{s}\|_2}{\|\mathbf{s}-\hat{\mathbf{s}}\|_2}, \quad (4)$$

where $\mathbf{s}$ is the ground truth and $\hat{\mathbf{s}}$ is the recovered signal.

2) **Perceptual Evaluation of Speech Quality (PESQ)**

PESQ is a popular measure for speech quality evaluation as experienced by a user of a telephony system. PESQ compares clean and noisy speech signals and returns a value between -0.5 and 4.5, with higher values indicating better quality.

3) **Short-Time Objective Intelligibility (STOI)**

STOI is a function that well represents the average intelligibility of the degraded speech. It provides a value from 0 to 1, with higher values indicating better intelligibility.

IV. RESULTS

*A. Changing the model architecture*

Table 2 shows the performance evaluation results of the baseline (MC Conv-TasNet) and the three models with their parameter sizes. The hyperparameters and parameter sizes of all models are presented in Tables 1 and 2, respectively. The size in the feature dimension ($F$) was fixed at 2048 for guaranteeing a reasonable comparison. In addition, the product of the channel and feature sizes for the TCN input was kept constant ($C \times N = 512$) to provide input tensors of the same total size irrespective of the models. The other parameters were determined using the values of a conventional Conv-TasNet [30], as presented in Table 2.

Table 2 illuminates that the 2-D Conv-TasNet significantly outperforms the baseline MC Conv-TasNet considering all performance indices. This improvement can be explained by the difference in encoder outputs. In the MC Conv-TasNet, the spatial information in the multichannel output from the encoder is lost by summing all channel outputs into a single-channel, whereas the 2-D Conv-TasNet preserves it by concatenating all encoder outputs.

The performance was further improved using the 3-D Conv-TasNet. The main difference between the 2-D and 3-D Conv-TasNets is the separation of the channel and feature dimensions. Because the channel-feature size product ($C \times N$) of the 3-D Conv-TasNet was equal to the total feature size ($N$) of the 2-D Conv-TasNet, the performance enhancement is solely attributable to the separation of the channel dimension. Unlike the simple concatenation of 2-D Conv-TasNet, the 3-D Conv-TasNet creates a new channel dimension by stacking the encoder outputs. Therefore, it is possible to maintain channel-dependent information through convolution operations. The significantly improved PESQ score of the 3-D Conv-TasNet implies that the separation of the channel dimension and 2-D dilated convolutions for time and feature dimensions help improve the sound quality.

Nevertheless, the 3-D Conv-TasNet has an intrinsic limitation in that different channel data are not mixed by the 1-D Conv layer. The TCN structure of the 3-D Conv-TasNet is merely a parallel connection of multiple TCNs for individual channel data. The inter-channel relationships are only utilized at the final mask generation step by the 1×1 Conv (*C*) layer.

This limitation of the 3-D Conv-TasNet can be resolved using the IC Conv-TasNet. According to the data presented in Table 2, the performance of IC Conv-TasNet surpasses all other models in every performance criterion. The IC Conv-TasNet has an improved TCN structure that aggregates the *C* channel data into *H* hidden layers, and subsequently the inter-channel relationships between them are extracted by the 1×1 Conv layers. Consequently, the IC Conv-TasNet makes use of spatial information more aggressively throughout the entire convolution process. Here, the parameter of IC Conv-TasNet that has the significant effect on the performance indices needs to be determined.

*B. Parameter study*

Parameter studies were conducted on the representative parameters summarized in Table 1 to investigate the sensitivity of the IC Conv-TasNet parameters. The variations in the model parameters are listed in Table 3 with the corresponding performance indices and total parameter sizes.

In the first part of the parameter study (Models 1–3), we examined the performance change with an increasing number of TCN stacks (*S*). The other model parameters were fixed as $D = 8$, $F = 2048$, $N = 64$, and $C = 8$. The performance gradually increases with an increasing *S*. However, when *S* is 3 or higher, the degree of improvement is not considerable. Although we can have a deeper TCN with a higher number of stacks, $S = 3$ was chosen for the next study regarding the increase in the parameter size in relationship to the change in *C* and *N*.

The performance can be further enhanced by increasing the number of dilated convolution layers *D*. When we changed *D* from 6 to 10 (Models 4, 1, and 5, respectively), an increase in SDR and STOI could be observed. However, the best PESQ was achieved for $D = 8$, which is partly because the effective receptive field size of the dilated convolution layer is limited to $2^8$ with a provided number of features ($N = 64$).

TABLE 3
PERFORMANCES OF IC CONV-TASNET MODELS WITH VARIOUS MODEL PARAMETERS

| Model number | D | S | F | N | C | H | SDR | PESQ | STOI | Param. size |
|---|---|---|---|---|---|---|---|---|---|---|
| 1 | 8 | 2 | 2048 | 64 | 8 | 32 | 16.29 | 2.15 | 0.949 | 1.34 M |
| 2 | 8 | 3 | 2048 | 64 | 8 | 32 | 16.51 | 2.24 | 0.949 | 1.35 M |
| 3 | 8 | 4 | 2048 | 64 | 8 | 32 | 16.85 | 2.24 | 0.953 | 1.36 M |
| 4 | 6 | 3 | 2048 | 64 | 8 | 32 | 15.90 | 2.14 | 0.947 | 1.34 M |
| 5 | 10 | 3 | 2048 | 64 | 8 | 32 | 16.54 | 2.17 | 0.951 | 1.35 M |
| 6 | 8 | 3 | 512 | 64 | 8 | 32 | 16.23 | 2.12 | 0.947 | 0.360 M |
| 7 | 8 | 3 | 512 | 128 | 8 | 32 | 17.28 | 2.33 | 0.960 | 0.425 M |
| 8 | 8 | 3 | 1024 | 128 | 8 | 32 | 17.42 | 2.28 | 0.959 | 0.820 M |
| 9 | 8 | 3 | 512 | 128 | 32 | 128 | 19.06 | 2.57 | 0.970 | 0.738 M |
| **10** | **8** | **3** | **512** | **128** | **64** | **256** | **19.67** | **2.67** | **0.973** | **1.67 M** |

In the next study, we investigated the influence of the number of encoder outputs ($F$). In Model 6 of Table 3, $F$ was decreased to 512 from 2048 in Model 1. Even with a 25% feature size, performance degradation is not considerable. The TCN features rather than the encoder outputs strongly influenced the performance of the proposed network. When the number of features in the output of the bottleneck layer ($N$), that is, the number of features for the TCN input, was increased to 128 from 64 of Model 6, noticeable improvements in SDR, PESQ, and STOI could be obtained (Model 7). This result reveals that the number of features considered in the dilated convolution layers is more important than the number of encoder outputs that are eventually downsized by the bottleneck layer. In addition, the significance of $N$ can be seen from the performance of Model 8, which has a higher $F$ than that of Model 7 with exhibiting no remarkable improvement.

From the insights gained in the aforementioned parameter studies, we configured a small-sized encoder output ($F = 512$) with a high number of features for the TCN input ($N = 128$). With this configuration, we continued the parameter study for the number of channels ($C$) in TCN. It was found that the number of channels is of paramount importance in enhancing the IC Conv-TasNet. Model 9, which has four times of $C$ than Model 7, yields dramatic improvements in SDR and PESQ. When $C$ was further increased to 64 (Model 10), a considerably higher performance could be obtained. These results demonstrate the importance of utilizing the channel and spatial information underlying multichannel speech data.

The spectrograms of a speech sample before and after enhancement by Model 10 are presented in Fig. 5. The spectrogram of the enhanced speech shows noise reduction over most frequency bands. Comparing Figs. 5 (b) and (c), however, the harmonics from 3–5 kHz are suppressed in enhanced speech compared to clean speech. This is attributable to the extremely low SNR of high frequency harmonics that cannot provide enough information for the reconstruction. Also, 1.7–2.1 kHz noise presents in the enhanced speech around 0.4 s, 1.1 s, and 3.7 s. Despite these observations, the enhanced speech quality is excellent and comparable to the clean speech in the informal listening test.

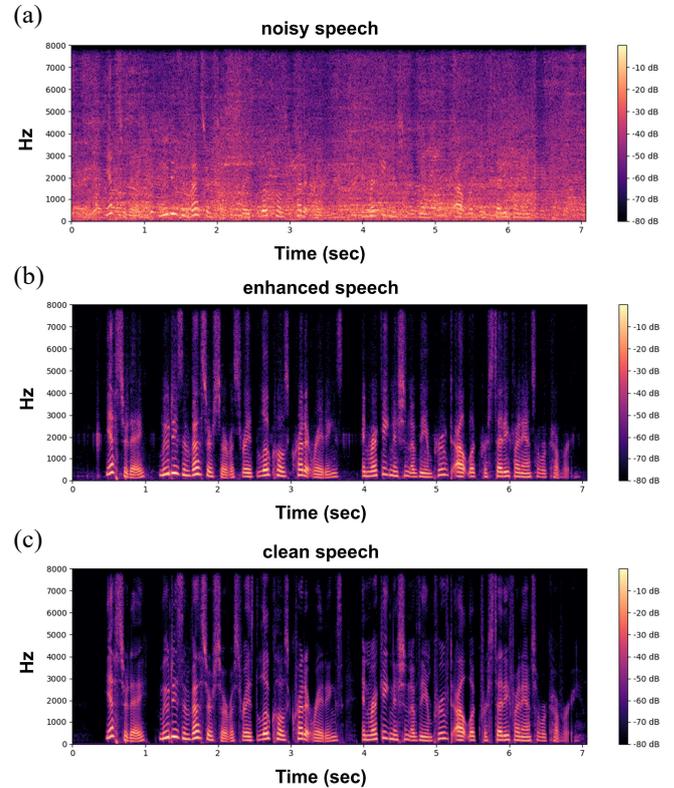

Fig. 5. Spectrogram of (a) noisy speech, (b) enhanced speech using IC Conv-TasNet, and (c) clean speech

Additionally, the proposed model performance was compared with that of the SOTA models [27], [37], which are based on different architectures for multichannel speech enhancement, and have the highest performance in SDR, PESQ, and STOI for the CHiME-3 dataset. According to the results summarized in Table 4, the proposed model exhibits an outstanding performance for all performance indices. Moreover, the improved performance can be obtained with a small parameter size for training. The parameter size of the proposed model is 1.67 M, which is only approximately 6.4% that of the CA Dense U-Net with the highest reported SDR performance. The small parameter size of the proposed model originates from the separation of the channel dimensions. In the conventional Conv-TasNet, the number of filters in the 1×1 Conv layers of TCN is determined by $N \times H = 128 \times 512$. Similarly, the number of filters in the D-Conv layer is $H = 512$. In contrast, in the IC Conv-TasNet, the number of filters of 1×1 Conv layers in TCN is $C \times H = 64 \times 256$, and that of D-Conv is $H = 256$. Therefore, the parameter size can be significantly reduced by separating the channel dimension.

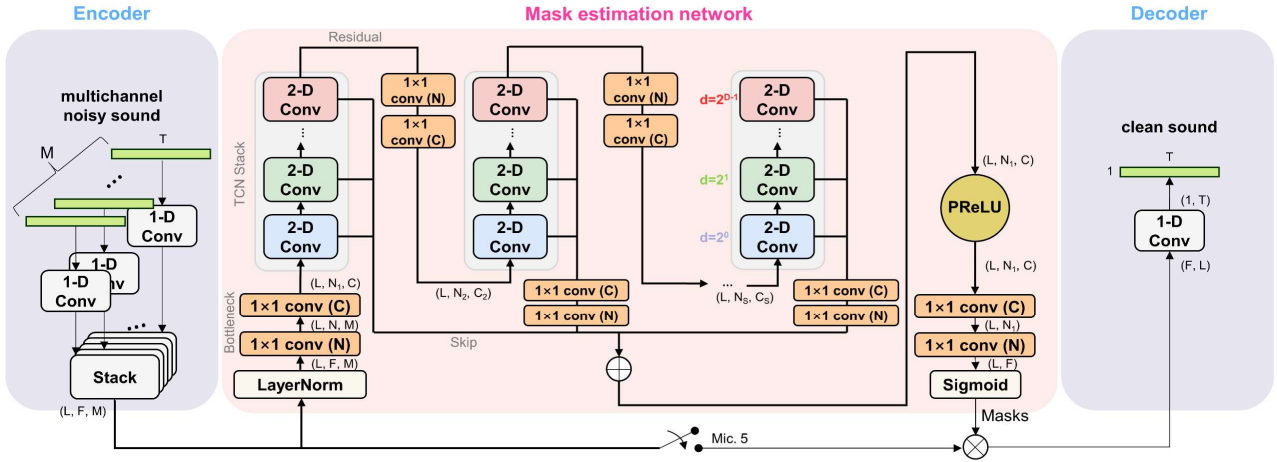

Fig. 6. Block diagram of overall structure of downsized IC Conv-TasNet

TABLE 4
PERFORMANCE COMPARISON OF IC CONV-TASNET WITH THAT OF THE STATE-OF-THE-ART MODELS

|  | SDR | PESQ | STOI | Param. Size |
|---|---|---|---|---|
| MVDR$_{GC}$ [37] | - | - | 0.952 | - |
| CA Dense U-Net [27] | 18.64 | 2.44 | - | 26 M |
| **IC Conv-TasNet** | **19.67** | **2.67** | **0.973** | **1.67 M** |

*C. Downsizing the IC Conv-TasNet*

Despite its small parameter size, training of the IC Conv-TasNet requires a considerable amount of GPU memory because size of the input and output tensors is greater than that of the conventional models. A downsized version of IC Conv-TasNet with smaller input tensors is developed (downsized IC Conv-TasNet) to reduce the memory consumption. The model architecture is shown in Fig. 6.

The downsized model (Model D) is characterized by a progressive output size reduction across individual TCN stacks. Since Model 10 shows the best performance among non-downsized models, Model D is based on Model 10. The input and output tensor sizes of the first TCN stack were the same as the previous case ($N_1 = N = 128$, $C_1 = C = 64$). However, as the stack index increases, the number of channels and features is progressively reduced. In the second stack, the number of channels is halved ($C_2 = C/2$), and the number of features is reduced to $N_2 = \lfloor N/\sqrt{2} \rfloor$. In the last ($S = 3$) stack, they are further reduced to $C_S = C/4$ and $N_S = N/2$. To superpose all skip connections from the 2-D Conv blocks, the size of each skip is increased to the original size of the first stack ($L, N, C$) in terms of two additional 1×1 Conv layers.

This progressive reduction structure is inspired by that of U-Net; however, we applied downsizing in channel and feature dimensions, whereas U-Net architecture pursuits the same in the time dimension. With the downsized model, sizes of the input and output tensors from each stack can be reduced by more than half compared with that of Model 10. This also reduces the parameter size and shortens the training time. In the experiment conducted with the same amount of GPU memory and machine, the training time was reduced by 70%.

Two other models were designed and compared with those of models D and 10. First, the upsizing model (Model U) increases from $(N_1, C_1) = (64, 16)$ to $(N_3, C_3) = (128, 64)$ as opposed to Model D. Sizes of the skip connections are also increased to match with the largest one (128, 64). Second, a small-sized IC Conv-TasNet model (Model S) of size $(N, C) = (64, 16)$ was examined without any progressive output size reduction to inspect the importance of progressive downsizing.

TABLE 5
PERFORMANCE COMPARISONS OF DOWNSIZED VERSIONS OF INTER-CHANNEL CONV-TASNET

|  | SDR | PESQ | STOI | Param. Size |
|---|---|---|---|---|
| **Model D** | **19.39** | **2.61** | **0.972** | **1.01 M** |
| Model U | 18.09 | 2.41 | 0.964 | 0.954 M |
| **Model 10** | **19.67** | **2.67** | **0.973** | **1.67 M** |
| Model S | 17.43 | 2.31 | 0.958 | 0.427 M |

In the results summarized in Table 5, Model 10 exhibits the highest performance; however, the downsized Model D also has a comparable performance. The performance of the downsized Model D is extremely higher than that of the small-sized Model

S; thus, it is necessary to have large-sized entry stacks in the progressive downsizing. This is in contrast to the upsizing Model U, which does not perform as well as Model D, implying that the size of the first stack is the most significant one, and loss of information in the first stack causes performance degradation in the following stacks.

## V. CONCLUSION

In this study, we proposed IC Conv-TasNet for a multichannel speech enhancement task. The proposed model introduced three major modifications to the conventional Conv-TasNet to effectively learn spatial information. First, in the encoder part, encoded multichannel signals were stacked along the new channel dimension to form a 3-D tensor, rather than summing to a single output channel. Therefore, the spatial information could be analyzed using the channel dimension, independent of the feature dimension. Second, the 1-D dilated convolution with respect to the feature dimension in TCN was changed to a 2-D dilated convolution for the feature and time dimensions. Thus, the channel-dependent information could be separately processed, which was implemented by 1×1 Conv layers that operated in the channel dimension rather than the feature dimension of the original Conv-TasNet. The performance of the proposed model was compared with that of the SOTA models tested using the same CHiME-3 dataset, which demonstrated excellent performance of the proposed model compared to that of the comparison models. In addition, a downsized IC Conv-TasNet model with small sizes of input and output tensors was proposed to reduce memory usage. For the speech enhancement task, the size of the first TCN stack was important; thus, the downsized model that progressively decreased the output tensor size with an increasing stack index exhibited a comparable performance to that of the full-sized IC Conv-TasNet.

## REFERENCES

[1] S. Yan, H. Sun, U. Svensson, X. Ma, and J. Hovem, "Optimal modal beamforming for spherical microphone arrays," *IEEE Trans. Audio, Speech, Lang. Process.*, vol. 19, no. 2, pp. 361–371, Feb. 2011.

[2] X. Li, S. Yan, X. Ma, and C. Hou, "Spherical harmonics MUSIC versus conventional MUSIC," *Applied Acoustics*, vol. 72, no. 9, pp. 646–652, Mar. 2011.

[3] Z. Zhao, H. Liu, and T. Fingscheidt, "Convolutional neural networks to enhance coded speech," *IEEE/ACM Trans. Audio, Speech, Lang. Process.*, vol. 27, no. 4, pp. 663–678, Apr. 2019.

[4] F. Weninger *et al.*, "Speech enhancement with LSTM recurrent neural networks and its application to noise-robust ASR," in *Proc. Int. Conf. LVA/ICA Liberec*, Czech Republic, 2015, pp. 91–99.

[5] T. Kounovsky and J. Malek, "Single channel speech enhancement using convolutional neural network," in *Proc. IEEE Int. Workshop ECMSM*, San Sebastian, Spain, 2017, pp. 1–5.

[6] X. Lu, Y. Tsao, S. Matsuda, and C. Hori, "Speech enhancement based on deep denoising autoencoder," in *Proc. Interspeech*, Lyon, France, 2013, pp. 436–440.

[7] S. Venkataramani, J. Casebeer, and P. Smaragdis, "End-to-end source separation with adaptive front-ends," in *Proc. 52nd Asilomar Conference on Signals, Systems, and Computers,* Pacific Grove, CA, USA, 2018, pp. 684–688.

[8] Y. Xu, J. Du, L. Dai, and C. Lee, "An experimental study on speech enhancement based on deep neural networks," *IEEE Signal processing letters*, vol. 21, no. 1, pp. 65–68, Nov. 2013.

[9] Y. Xu, J. Du, L. Dai, and C. Lee, "A regression approach to speech enhancement based on deep neural networks." *IEEE/ACM Trans. Audio, Speech, Lang. Process.*, vol. 23, no. 1, pp. 7–19, Jan. 2015.

[10] S. R. Park and J. Lee, "A fully convolutional neural network for speech enhancement", in *Proc. Interspeech*, Stockholm, Sweden, 2017, pp. 1993-1997.

[11] D. S. Williamson, Y. Wang, and D. Wang, "Complex ratio masking for monaural speech separation," *IEEE/ACM Trans. Audio, Speech, Lang. Process.* vol. 24, no. 3, pp. 483–492, Dec. 2015.

[12] D. Wang and J. Chen, "Supervised speech separation based on deep learning: An overview," *IEEE/ACM Trans. Audio, Speech, Lang. Process.*, vol. 26, no. 10, pp. 1702–1726, May 2018.

[13] S. Fu, Y. Tsao, X. Lu, and H. Kawai, "Raw waveform-based speech enhancement by fully convolutional networks," in *Proc. Asia-Pacific Signal and Information Process. Assoc. Annu. Summit and Conf. (APSIPA ASC)*, Kuala Lumpur, Malaysia, 2017, pp. 6–12.

[14] A. Narayanan and D. Wang, "Ideal ratio mask estimation using deep neural networks for robust speech recognition," in *Proc. IEEE Int. Conf. Acoust., Speech, Signal Process.*, Vancouver, BC, Canada, 2013, pp. 7092–7096.

[15] G. Kim, Y. Lu, Y. Hu, and P. Loizou, "An algorithm that improves speech intelligibility in noise for normal-hearing listeners," *J. Acoust. Soc. Amer.*, vol. 126, no. 3, pp. 1486–1494, Sep. 2009.

[16] D. Wang, "On ideal binary mask as the computational goal of auditory scene analysis," in *Speech separation by humans and machines*, Boston, MA, USA: Springer, 2005, pp. 187–197.

[17] D. Wang and G. Brown, *Computational auditory scene analysis: Principles, algorithms, and applications*. in Hoboken, NJ, USA, Wiley, 2006.

[18] Y. Xu, J. Du, L. Dai, and C. Lee, "A regression approach to speech enhancement based on deep neural networks," *IEEE/ACM Trans. Audio, Speech, Lang. Process.*, vol 23, no. 1, pp. 7–19, Oct. 2014.

[19] Y. Li and D. Wang, "On the optimality of ideal binary time-frequency masks," *Speech Communication*, vol. 51, no. 3, pp. 230–239, Mar. 2009.

[20] K. Tan and D. Wang, "A convolutional recurrent neural network for real-time speech enhancement," in *Proc. Interspeech*, Hyderabad, India, 2018, pp. 3229–3233.

[21] Y. Luo and N. Mesgarani, "Tasnet: time-domain audio separation network for real-time, single-channel speech separation," in *Proc. IEEE Int. Conf. Acoust., Speech, Signal Process.*, Calgary, Alberta, Canada, 2018, pp. 696–700.

[22] H. Erdogan, J. R. Hershey, S. Watanabe, M. Mandel, and J. Le Roux, "Improved mvdr beamforming using single-channel mask prediction networks," in *Proc. Interspeech*, San Francisco, CA, USA, 2016, pp. 1981–1985.

[23] J. Heyman, L. Drude, and R. Haeb-Umbach, "Neural network based spectral mask estimation for acoustic beamforming," in *Proc. IEEE Int. Conf. Acoust., Speech, Signal Process.*, Shanghai, China, 2016, pp. 196–200.

[24] Y. Liu, A. Ganguly, K. Kamath, and T. Kristjansson, "Neural network based time-frequency masking and steering vector estimation for two-channel mvdr beamforming", in *Proc. IEEE Int. Conf. Acoust., Speech, Signal Process.*, Calgary, Alberta, Canada, 2018, pp. 6717–6721.

[25] K. Shimada, Y. Bando, M. Mimura, K. Itoyama, K. Yoshii, and T. Kawahara, "Unsupervised speech enhancement based on multichannel NMF-informed beamforming for noise-robust automatic speech recognition," *IEEE/ACM Trans. Audio, Speech, Lang. Process.*, vol. 27, no. 5, pp.960–971, May 2019.

[26] M. Togami, "Simultaneous optimization of forgetting factor and time-frequency mask for block online multi-channel speech enhancement," in *Proc. IEEE Int. Conf. Acoust., Speech, Signal Process.*, Brighton, UK, 2019, pp. 2702–2706.

[27] B. Tolooshams, R. Giri, A. Song, U. Isik, and A. Krishnaswamy, "Channel-attention dense u-net for multichannel speech enhancement," in *Proc. IEEE Int. Conf. Acoust., Speech, Signal Process.*, Barcelona, Spain, 2020, pp.836–840.

[28] J. Barker, R. Marxer, E. Vincent and S. Watanabe, "The third 'CHiME' speech separation and recognition challenge: Dataset, task and baselines," in *IEEE Workshop Autom. Speech Recognit. Understanding*, Scottsdale, AZ, USA, 2015, pp. 504–511.

[29] O. Ronneberger, P. Fischer, and T. Brox, "U-Net: Convolutional Networks for Biomedical Image Segmentation," in *Proc. Med. Image Comput. Comput.-Assisted Intervention*, Munich, Germany, 2015, pp. 234–241.

[30] Y. Luo and N. Mesgarani, "Conv-TasNet: Surpassing ideal time-frequency magnitude masking for speech separation," *IEEE/ACM Trans. Audio, Speech, Lang. Process.*, vol. 27, no. 8, pp. 1256–1266, Aug, 2019.

[31] C. Lea, R. Vidal, A. Reiter, and G. Hager, "Temporal convolutional networks: A unified approach to action segmentation," in *Eur. Conf. Comput. Vis.*, Amsterdam, Netherland, 2016, pp. 47–54.

[32] C. Lea, M. Flynn, R. Vidal, A. Reiter, and G. Hager, "Temporal convolutional networks for action segmentation and detection," in *Proc. IEEE Conf. Comput. Vis. Pattern Recognit.*, Honolulu, Hawaii, USA, 2017, pp. 156–165.

[33] Y. Koyama, T. Vuong, S. Uhlich, and B. Raj, "Exploring the best loss function for DNN-based low-latency speech enhancement with temporal convolutional networks," 2020, *arXiv:2005.11611*.

[34] F. Xiao, J. Guan, Q. Kong, and W. Wang, "Time-domain speech enhancement with generative adversarial learning," 2021, *arXiv:2103.16149*.

[35] R. Gu *et al.*, "End-to-end multi-channel speech separation," 2019, *arXiv:1905.06286*.

[36] R. Gu *et al.*, "Enhancing end-to-end multi-channel speech separation via spatial feature learning," in *Proc. IEEE Int. Conf. Acoust., Speech, Signal Process.*, Barcelona, Spain, 2020. pp. 7319–7323.

[37] S. Bu, Y. Zhao, and M. Hwang, "A novel method to correct steering vectors in MVDR beamformer for noise robust ASR," in *Proc. Interspeech*, Graz, Austria, 2019, pp. 4280–4284.

[38] A. Agarap, "Deep learning using rectified linear units (relu)," 2018, *arXiv:1803.08375*.

[39] C. Deng, Y. Zhang, S. Ma, Y. Sha, H. Song, and X. Li, "Conv-TasSAN: Separative adversarial network based on Conv-TasNet," in *Proc. Interspeech*, Shanghai, China, 2020, pp. 2647–2651.

[40] F. Chollet, "Xception: Deep learning with depthwise separable convolutions," in *Proc. IEEE Conf. Comput. Vis. Pattern Recognit.*, Honolulu, Hawaii, USA, 2017, pp. 1251–1258.

[41] A. G. Howard *et al.*, "Mobilenets: Efficient convolutional neural networks for mobile vision applications," 2017, *arXiv:1704.04861*.

[42] K. He, X. Zhang, S. Ren, and J. Sun, "Delving deep into rectifiers: Surpassing human-level performance on imagenet classification," in *Proc. IEEE Int. Conf. Comput. Vis.*, Santiago, Chile, 2015, pp. 1026–1034.

[43] T. Vu, B. Bigot, and E. Chng, "Speech enhancement using beamforming and non negative matrix factorization for robust speech recognition in the CHiME-3 challenge," in *IEEE Workshop Autom. Speech Recognit. Understanding*, Scottsdale, AZ, USA, 2015, pp. 423–429.

[44] D. P. Kingma and J. Ba, "Adam: A method for stochastic optimization," 2014, *arXiv:1412.6980*.

[45] E. Vincent, R. Gribonval, and C. Févotte, "Performance measurement in blind audio source separation," *IEEE Trans. Audio, Speech, Lang. Process.*, vol. 14, no. 4, pp. 1462–1469, Jun. 2006.

[46] A. Rix, J. Beerends, M. Hollier, and A. Hekstra, "Perceptual evaluation of speech quality (PESQ)-a new method for speech quality assessment of telephone networks and codecs," in *Proc. IEEE Int. Conf. Acoust. Speech, Signal Process.*, Salt Lake City, UT, USA, 2001, pp. 749–752.

[47] C. Taal, R. Hendriks, R. Heusdens, and J. Jensen, "An algorithm for intelligibility prediction of time-frequency weighted noisy speech," *IEEE Trans. Audio, Speech, Lang. Process.*, vol. 19, no. 7, pp. 2125–2136, Sep. 2011.